\documentclass[aps,prd,showpacs,nofootinbib,onecolumn,widetext,amsmath,amssymb]{revtex4} 
\usepackage{latexsym}
\usepackage{amsfonts}
\usepackage{amsbsy}
\usepackage{mathrsfs}

\newcommand{\h}{\tilde{H}_{\mu\nu}}
\newcommand{\g}{\tilde{G}_{\mu\nu}}
\newcommand{\n}{\,\tilde{\eta}^{\alpha\beta\gamma\delta}}
\newcommand{\dete}{\det(e_\alpha^I)}

\newcommand{\ff}{\mathcal{F}}
\newcommand{\A}{\left(\frac{1}{\beta^2_2-\beta_1^{2}\sigma}\right)}

\begin{document}

\title{BF gravity with Immirzi parameter and matter fields}

\date{\today}

\author{Merced Montesinos}\email{merced@fis.cinvestav.mx}

\author{Mercedes Vel\'azquez}\email{mquesada@fis.cinvestav.mx}

\affiliation{Departamento de F\'{\i}sica, Cinvestav, Instituto Polit\'ecnico
Nacional 2508, San Pedro Zacatenco, 07360, Gustavo A. Madero, Ciudad de
M\'exico, M\'exico.}

\begin{abstract}
We perform the coupling of the scalar, Maxwell, and Yang-Mills fields as well as the cosmological constant to $BF$
gravity with Immirzi parameter. The proposed action principles employ auxiliary fields in order to keep a polynomial
dependence on the $B$ fields. By handling the equations of motion for the $B$ field and for the auxiliary fields, these
latter can be expressed in terms of the physical fields and by substituting these expressions into the original action
principles we recover the first-order (Holst) and second-order actions for gravity coupled to the physical matter
fields. We consider these results a relevant step towards the understanding of the coupling of matter fields to gravity
in the theoretical framework of $BF$ theory.
\end{abstract}


\maketitle
\section{Introduction}

The research in quantum gravity led by its two main branches (loop quantum gravity \cite{lqg} and spin foam models for
gravity \cite{spinfoam}) has recently motivated the study of the classical descriptions for general relativity and
theories related to it, particularly the formulations of gravity as a constrained BF theory. Just to mention some of
them, Cartan's equations in the framework of BF theories are analyzed in Refs. \onlinecite{vlad1} and \onlinecite{vlad2} while the
relationship of general relativity to the Husain-Kuchar model in the framework of BF theory is analyzed in Refs.
\onlinecite{meche1,meche2,cuando}.

It is possible to say that loop quantum gravity and spin foam models for gravity are, in a certain sense, inspired in
the Plebla\'nski's work \cite{pleb77}. As is well-known, in the mid-70's of the twentieth century Pleba\'nski wrote the
equations of motion for four-dimensional general relativity in such a way that the fundamental variables for describing
the gravitational field are two-form fields, a connection one-form, and some Lagrange multipliers. The geometry of the
spacetime is built up from these fundamental blocks. The Pleba\'nski action is a $BF$ theory supplemented with
constraints on some of the fields involved. In order to bring tetrads into the formulation, the two-forms $B$'s are
eliminated by solving the constraints on them, which implies that the $B$'s can be expressed in terms of tetrad fields,
and by inserting back this into Pleba\'nski's action, it becomes the self-dual action for general relativity
\cite{1Samuel,1Jacobson}. At the beginning of the 1990's Pleba\'nski's formulation was extended in order to include the
coupling of matter fields \cite{cdjm}.

Following the same idea used by Pleba\'nski, at the beginning of this century an action principle for real general
relativity including the Immirzi parameter was introduced by Capovilla, Montesinos, Prieto and Rojas in Ref.
\onlinecite{cqgl2001} (CMPR action principle). This action is very close to the Pleba\'nski action because it is given
in terms of two-forms, a connection one-form, and some Lagrange multipliers but now it includes an arbitrary value of
the Immirzi parameter. This is not the unique formulation for real $BF$ gravity, there also exist another one used by
Engle, Pereira, and Rovelli in Ref. \onlinecite{epr}. These two formulations can be related by doing a transformation
between the set of fields involved in one of them and the set of fields involved in the other. A detailed analysis of
this transformation can be found in Ref. \onlinecite{sigma1} (see also Ref. \onlinecite{oriti}). The transformation
allows us to translate any analysis made in one formulation to the other one. In particular, the coupling of matter
fields to $BF$ gravity with Immirzi parameter can be done in any of these two approaches, and using the transformation
such a coupling can be done in the other framework. As an example of this theoretical framework, the inclusion of the
cosmological constant to the action principle used in Ref. \onlinecite{epr} was obtained in Ref. \onlinecite{sigma1}
from the coupling of the cosmological constant to the CMPR action principle done in Ref. \onlinecite{prd2010}.

Following this trend of ideas, in this paper we focus our attention in the coupling of the cosmological constant, the
scalar, Yang-Mills, and Maxwell fields to gravity in the $BF$ framework. Our starting point is the action principle
used in Ref. \onlinecite{epr} but, as mentioned above, once the coupling is done, it is also possible to obtain such a
coupling in the CMPR action principle. For each one of the couplings presented in this paper, it can be shown that
Einstein's equations and the equations of motion corresponding to the involved matter field follow immediately from the
proposed action principle. This will be explicitly shown for the coupling of the cosmological constant while for the
coupling of the other matter fields, the proposed action principles will be rewritten in terms of tetrad fields by
solving the constraints on the two-forms and the equations of motion for the auxiliary fields. Following this approach,
we will obtain Holst's action principle for gravity coupled with the correspondent matter field. Owing to this
equivalence, it is pretty obvious that the equations of motion that follow from the proposed action principles are
equivalent to Einstein's equations coupled to matter fields plus the equations of motion for the matter fields
themselves. The material reported in this paper is part of the work presented in Ref. \onlinecite{mpvq2}.

The starting point for the coupling of matter fields to gravity is the action principle for general relativity given by
\begin{eqnarray}\label{eprl}
S_{GR}[B,A,\Phi,\mu]=\int_{\mathcal{M}^4} \left [ \left ( B^{IJ} + \frac{1}{\gamma}\, {^{\ast} B}^{IJ} \right ) \wedge
F_{IJ} [A] - \frac12 \Phi_{IJKL} B^{IJ} \wedge B^{KL}
 - \mu \,  \Phi_{IJKL}\, \varepsilon^{IJKL} \right ],
\end{eqnarray}
where $F^I\,_J[A]= d A^I\,_{J} + A^I\,_{K} \wedge A^K\,_J$ is the curvature of the $SO(4)$ or $SO(3,1)$ connection
one-form $A^{IJ}=-A^{JI}$ and the $B^{IJ}$ are six two-forms because of the property $B^{IJ}=-B^{JI}$; the internal
indices $I,J,K,\ldots$ take on the values 0,1,2,3 and they are raised and lowered with the metric
$(\eta_{IJ})=(\sigma,1,1,1)$ where $\sigma=-1$ for Lorentzian and $+1$ for Euclidean signatures, respectively. We
define $^{*}B^{IJ}=\frac12\varepsilon^{IJ}\,_{KL}B^{KL}$, $\mu$ is a 4-form and $\Phi_{IJKL}$ has the usual symmetries
$\Phi_{IJKL}=\Phi_{KLIJ}$, $\Phi_{IJKL}=-\Phi_{JIKL}$, and $\Phi_{IJKL}=-\Phi_{IJLK}$. See Refs. \onlinecite{celada}
and \onlinecite{celada2} for the Lorentz-covariant Hamiltonian analysis of the action (\ref{eprl}).

\section{Coupling the cosmological constant}\label{cc}

The coupling of the cosmological constant to the CMPR action principle has been studied in Refs. \onlinecite{prd2010}
and \onlinecite{simone}. The coupling of the cosmological constant presented here is done following our ideas reported
in Refs. \onlinecite{prd2010} and \onlinecite{merced}. In order to introduce the cosmological constant into the action
principle (\ref{eprl}), we propose the following action principle
\begin{eqnarray}\label{eprl+cc}
S[B,A,\Phi,\mu] =S_{GR}[B,A,\Phi,\mu]+ \int_{\mathscr{M}^4} \left ( \mu \lambda + l_1 B_{IJ}\wedge B^{IJ} + l_2
B_{IJ}\wedge \,^{\ast}B^{IJ}\right ),
\end{eqnarray}
where $\lambda$, $l_1$, and $l_2$ are constants whose relationship with the cosmological constant $\Lambda$ will be
analyzed below. The variation of this action gives the equations of motion
\begin{subequations}\label{eq-CC}
\begin{eqnarray}
&& \delta B: F_{IJ} [A] + \frac{1}{\gamma} \,^{\ast}F_{IJ}-
\Phi_{IJKL} B^{KL} + 2l_1 B_{IJ} + 2l_2\, ^{\ast}B_{IJ}=0, \label{eq B CC}\\
&& \delta A: D B^{IJ} + \frac{1}{\gamma} \,D\,^{\ast}B^{IJ} =0, \label{eq A CC}\\
&& \delta \Phi: B^{IJ} \wedge B^{KL} + 2 \mu\, \varepsilon^{IJKL}=0,\label{eq Phi CC}\\
&& \delta \mu:  \Phi_{IJKL} \varepsilon^{IJKL} =\lambda. \label{eq mu CC}
\end{eqnarray}
\end{subequations}
In what follows it will be shown that Eqs. (\ref{eq-CC}) imply that the Einstein's equations with cosmological constat
given by
\begin{eqnarray}\label{einstein+cc}
{^{\ast}F}_{IJKL} + {^{\ast}F}_{IKLJ} + {^{\ast}F}_{ILJK}= \Lambda \varepsilon_{IJKL},
\end{eqnarray}
with $F_{IJ}=\frac12 F_{IJKL} e^K \wedge e^L$, are completely satisfied.

We are going to begin the analysis of the equations of motion (\ref{eq-CC}). It can be easily seen that Eq. (\ref{eq B
CC}) can be written as\footnote{Note that the fact that Eq. (\ref{forma F CC}) holds for $\gamma^2 \neq \sigma$ is a
feature already present in the action (\ref{eprl}) for pure gravity and does not come from the coupling of the
cosmological constant itself. Furthermore, the action (\ref{eprl}) does {\it not} reduce to the Pleba\'nski action
\cite{pleb77} for the choices of $\gamma^2 =\sigma$ as it is explained in detail in Ref. \onlinecite{mpvq2} where, by
the way, an action principle having the right self-dual limits is reported. In spite of this property, the action
(\ref{eprl}) has been subject of study by Engle, Pereira, and Rovelli in Ref. \onlinecite{epr} and more recently by
Perez in the last paper of Ref. \onlinecite{spinfoam}.}
\begin{eqnarray}\label{forma F CC}
F_{IJ}=\frac{\gamma^2}{\gamma^2-\sigma}\left[ \left( \Phi_{IJKL}-\frac{1}{\gamma}\,^{\ast}\Phi_{IJKL}\right)B^{KL}
+2\left(\frac{l_2 \sigma}{\gamma}-l_1 \right)B_{IJ} +2\left(\frac{l_1}{\gamma}-l_2 \right)\,^{\ast}B_{IJ} \right],
\end{eqnarray}
provided that $\gamma^2\neq\sigma$. Under the same restriction, Eq. (\ref{eq A CC}) reduces to
\begin{eqnarray}\label{DB=0}
DB^{IJ}=0.
\end{eqnarray}
On the other hand, Eqs. (\ref{eq Phi CC}) imply that there exist two independent solutions for the two-forms $B^{IJ}$
given by
\begin{subequations}\label{sol B's}
 \begin{eqnarray}\label{sol B's1}
\quad B^{IJ}&=& \kappa_1 \, {^{\ast}} \left ( e^I \wedge e^J \right ),\\
\quad B^{IJ}&=& \kappa _2\, e^I \wedge e^J, \label{sol B's2}
 \end{eqnarray}
\end{subequations} where $\kappa_1$ and $\kappa_2$ are constants. For any of these solutions, Eq. (\ref{DB=0})
reduces to $D e^{I}=0$. This means that $A^{I}\,_{J}$ is the spin-connection. Therefore, the curvature $F^{I}\,_{J}$
must satisfy the Bianchi identities without torsion given by
\begin{eqnarray}\label{bianchi}
F_{IJKL} + F_{IKLJ} + F_{ILJK}=0.
\end{eqnarray}
In summary, Eqs. (\ref{eq A CC})  and (\ref{eq Phi CC}) imply that $A^I\,_J$ is the spin-connection, $F^{I}\,_{J}$
satisfies Bianchi identities without torsion (\ref{bianchi}), and that the two-form $B$ can be written as given in
(\ref{sol B's}). Moreover, due to the fact that Eq. (\ref{forma F CC}) expresses the curvature $F_{IJ}$ in  terms of
the Lagrange multiplier $\Phi_{IJKL}$ and the $B$ field, it is clear that $\Phi_{IJKL}$ and $B^{IJ}$ must satisfy some
restrictions coming from the fulfilling of the Bianchi identities. In addition, the substitution of the solutions given
in Eq. (\ref{sol B's}) into Eq. (\ref{forma F CC}) will give two different expressions for the curvature in terms of
the tetrad field $e^I$. Let us consider each case separately.

\subsection{Case $B^{IJ}=\kappa_1 \, {^{\ast}} \left ( e^I \wedge e^J \right )$}\label{cc c1}
Substituting the two-form (\ref{sol B's1}) into the Eq. (\ref{forma F CC}) the components of the curvature take the
form
\begin{eqnarray}\label{f c1}
F_{IJKL}=\frac{2 \gamma^2 \kappa_1}{\gamma^2-\sigma} \left[ \Phi^{\ast}_{IJKL} -
\frac{1}{\gamma}\,^{\ast}\Phi^{\ast}_{IJKL} +\left(\frac{l_2 \sigma}{\gamma}-l_1 \right)\varepsilon_{IJKL} +2\sigma
\left(\frac{l_1}{\gamma}-l_2 \right) \eta_{[I\mid K \mid} \eta_{J]L}\right].
\end{eqnarray}

Inserting (\ref{f c1}) into the Bianchi identities (\ref{bianchi}), we get the next relations among the components of
the field $\Phi_{IJKL}$
\begin{eqnarray}\label{rest c1}
\Phi^{\ast}_{IJKL} + \Phi^{\ast}_{IKLJ} +\Phi^{\ast}_{ILJK} -\frac{1}{\gamma} \left( \,^{\ast}\Phi^{\ast}_{IJKL} +
\,^{\ast}\Phi^{\ast}_{IKLJ} +\,^{\ast}\Phi^{\ast}_{ILJK}\right) +3\left(\frac{l_2 \sigma}{\gamma}-l_1
\right)\varepsilon_{IJKL}=0.
\end{eqnarray}

According to Ref. \onlinecite{prd2010}, the next step is to introduce (\ref{f c1}) into the Einstein's equations with
cosmological constant (\ref{einstein+cc}) and use the restrictions (\ref{rest c1}) in order to check whether the
Einstein's equations with cosmological constant are satisfied or not. It is obtained that the solely fulfilling of the
Bianchi identities implies the fulfilling of (\ref{einstein+cc}), except by one equation given by
\begin{eqnarray}\label{E4 c1}
\Phi_{IJ}\,^{IJ}-\frac{\gamma \sigma}{2} \Phi_{IJKL} \varepsilon^{IJKL} =
-\frac{2\Lambda\sigma}{\kappa_1}\left(\frac{\gamma^2-\sigma}{\gamma}\right) +12\left(l_1-l_2\gamma \right),
\end{eqnarray}
whose left-hand side involves a linear combination of the two Lorentz invariants $\Phi^{IJ}\,_{IJ}$ and
$\Phi_{IJKL}\varepsilon^{IJKL}$. Therefore, the remaining task is to be sure that (\ref{E4 c1}) comes effectively from
(\ref{eq-CC}). To get this goal, we note that other equation that relates the two Lorentz invariants $\Phi^{IJ}\,_{IJ}$
and $\Phi_{IJKL}\varepsilon^{IJKL}$ comes from the Bianchi identities and is obtained by contracting (\ref{rest c1})
with $\varepsilon^{IJKL}$, which leads to
\begin{eqnarray}\label{B4 c1}
\Phi_{IJ}\,^{IJ}-\frac1{2\gamma} \Phi_{IJKL} \varepsilon^{IJKL} = -12\left(\frac{l_2 \sigma}{\gamma}-l_1 \right).
\end{eqnarray}
However, (\ref{B4 c1}) is not enough to satisfy (\ref{E4 c1}). Nevertheless, the combination of (\ref{B4 c1}) with the
equation of motion (\ref{eq mu CC}) yields to
\begin{eqnarray}\label{B4+em mu c1}
\Phi_{IJ}\,^{IJ}-\frac{\gamma \sigma}{2} \Phi_{IJKL} \varepsilon^{IJKL} = -12 \left(\frac{l_2 \sigma}{\gamma}-l_1
\right) -\frac{\lambda\sigma}{2}\left(\frac{\gamma^2-\sigma}{\gamma}\right).
\end{eqnarray}
Comparing (\ref{E4 c1}) with (\ref{B4+em mu c1}) we see that their right-hand-sides are equal to each other provided
that
\begin{eqnarray}\label{H c1}
\lambda= \frac{4\Lambda}{\kappa_1}+4! l_2 \sigma.
\end{eqnarray}
This means that the equations of motion obtained from the action principle (\ref{eprl+cc}) with the value of $\lambda$
given in (\ref{H c1}), imply that the Einstein's equations with cosmological constant are completely satisfied when Eq.
(\ref{eq Phi CC}) is solved by (\ref{sol B's1}). Note that (\ref{H c1}) relates the constant $\kappa_1$ of the solution
(\ref{sol B's1}) with the value of $\lambda$ in the action principle.

Alternatively, it is possible to write the action principle (\ref{eprl+cc}) in the usual and equivalent form given by
Holst's action for general relativity with cosmological constant
\begin{eqnarray}\label{epr+cc en e's c1}
S[e,A]=\kappa_1 \int_{{\mathscr M}^4}\left\{\left[{^{\ast}} \left( e^I \wedge e^J \right) +\frac{\sigma}{\gamma}\, e^I
\wedge e^J \right]\wedge F_{IJ}[A]-\frac{\Lambda}{12}\varepsilon_{IJKL} e^I\wedge e^J\wedge e^K\wedge e^L\right\},
\end{eqnarray}
which is obtained by substituting the expression (\ref{sol B's1}) into (\ref{eprl+cc}) together with the value for
$\lambda$ given in (\ref{H c1}) and the value of $\mu$ obtained from (\ref{eq Phi CC}).

\subsection{Case $B^{IJ}=\kappa_2\,  e^I \wedge e^J $}\label{cc c2}

In this case the introduction of (\ref{sol B's2}) into Eq. (\ref{forma F CC}) gives for the curvature
\begin{eqnarray}\label{f c2}
F_{IJKL}=\frac{2 \gamma^2 \kappa_2}{\gamma^2-\sigma} \left[ \Phi_{IJKL} - \frac{1}{\gamma}\,^{\ast}\Phi_{IJKL}
+2\left(\frac{l_2\sigma}{\gamma}-l_1 \right) \eta_{[I\mid K \mid} \eta_{J]L} + \left(\frac{l_1}{\gamma}-l_2 \right)
\varepsilon_{IJKL} \right].
\end{eqnarray}
Once again, by inserting (\ref{f c2}) into the Bianchi identities (\ref{bianchi}) yields to the restrictions on the
$\Phi_{IJKL}$ field
\begin{eqnarray}\label{rest c2}
\Phi_{IJKL} + \Phi_{IKLJ} +\Phi_{ILJK} -\frac{1}{\gamma} \left( \,^{\ast}\Phi_{IJKL} + \,^{\ast}\Phi_{IKLJ}
+\,^{\ast}\Phi_{ILJK}\right) +3 \left(\frac{l_1}{\gamma}-l_2 \right) \varepsilon_{IJKL}=0.
\end{eqnarray}
The contraction of (\ref{rest c2}) with $\varepsilon^{IJKL}$ gives the equation for the two Lorentz invariants
\begin{eqnarray}\label{B4 c2}
\Phi_{IJ}\,^{IJ}-\frac{\gamma\sigma}2 \Phi_{IJKL} \varepsilon^{IJKL} = 12\left(l_1-l_2 \gamma \right),
\end{eqnarray}
which can be combined with the equation of motion (\ref{eq mu CC}) to give
\begin{eqnarray}\label{B4+em mu c2}
\Phi_{IJ}\,^{IJ}-\frac1{2\gamma} \Phi_{IJKL} \varepsilon^{IJKL} = 12\left(l_1-l_2 \gamma \right)
+\frac{\lambda\sigma}{2} \left(\frac{\gamma^2-\sigma}{\gamma}\right).
\end{eqnarray}
On the other hand, using the form of the curvature given in (\ref{f c2}) and the restrictions (\ref{rest c2}) in the
Einstein's equations (\ref{einstein+cc}) it is concluded that they all are automatically satisfied, except by the
equation given by
\begin{eqnarray}\label{E4 c2}
\Phi_{IJ}\,^{IJ}-\frac1{2\gamma} \Phi_{IJKL} \varepsilon^{IJKL} =
\frac{2\Lambda}{\kappa_2}\left(\frac{\gamma^2-\sigma}{\gamma^2}\right) -12\left(\frac{l_2\sigma}{\gamma} - l_1 \right).
\end{eqnarray}
Comparing the right-hand-sides of (\ref{B4+em mu c2}) and (\ref{E4 c2}) we fix the value for $\lambda$ to
\begin{eqnarray}\label{H c2}
\lambda= \sigma \left(\frac{4\Lambda}{\kappa_2\gamma}+4!\, l_2 \right).
\end{eqnarray}
As it happens in Sec. \ref{cc c1}, this means that the equations of motion obtained from the action principle
(\ref{eprl+cc}) with the value of $\lambda$ given in (\ref{H c2}) imply that the Einstein's equations with cosmological
constant are totally satisfied when Eq. (\ref{eq Phi CC}) is solved by (\ref{sol B's2}). Note that (\ref{H c2}) relates
the constant $\kappa_2$ of the solution with the value of $\lambda$ in the action principle.

Again, if the expression (\ref{sol B's2}) is substituted into (\ref{eprl+cc}) together with the value for $\lambda$
given in (\ref{H c2}) and the value of $\mu$ obtained from (\ref{eq Phi CC}), it is possible to write the action
principle (\ref{eprl+cc}) in terms of the tetrad field an a Lorentz connection as
\begin{eqnarray}
S[e,A]=\frac{\kappa_2}{\gamma} \int_{{\mathscr M}^4}\left\{\left[ \,{^{\ast}}\left(e^I \wedge e^J\right) + \gamma\,
 e^I \wedge e^J  \right]\wedge F_{IJ}[A] -\frac{\Lambda}{12}\varepsilon_{IJKL} e^I\wedge e^J\wedge e^K\wedge e^L\right\},
\end{eqnarray}
which is of the form of the Holst action, as the one obtained in section \ref{cc c1}, but with a different expression
for the Immirzi parameter (confront with Eq. (\ref{epr+cc en e's c1})).

It should be noticed that there is a subtle difference between the two cases A and B previously discussed. Even though
in the two cases we have the coupling of the cosmological constant, in the case B Newton's constant involves a $\gamma$
factor.

We conclude by remarking that the constant $l_1$ does not appear in any of the two values obtained for $\lambda$ in
Eqs. (\ref{H c1}) and (\ref{H c2}). This is because even though in the action principle (2) the two allowed volume
terms \cite{prd2010,merced}, $l_1 B_{IJ}\wedge B^{IJ}$ and $l_2 B_{IJ}\wedge \,^{\ast}B^{IJ}$, are included, from Eq.
(\ref{eq Phi CC}) it can be seen that $B_{IJ}\wedge B^{IJ}=0$ for any solution of the $B$'s, thus when the action
(\ref{eprl+cc}) is written in terms of the tetrad field, the term $l_1 B_{IJ}\wedge B^{IJ}$ identically vanishes.
Nevertheless the inclusion of such a term into the action principle (\ref{eprl+cc}) does affect the value of the
Lorentz invariant $\Phi_{IJ}\,^{IJ}$ (see Eqs. (\ref{B4+em mu c1}) and (\ref{B4+em mu c2})).

\section{Coupling the scalar field}

Continuing with the analysis now we consider the coupling of a scalar field. The action principle (\ref{eprl}) (or the
one considered in Ref. \onlinecite{cqgl2001}) is quadratic in the $B$ fields. Therefore, it is natural to keep in the action
principle a polynomial dependence on the $B$'s when the coupling of a scalar field $\phi$ is done. This can be achieved
by introducing auxiliary fields $\pi^{\mu}$ \cite{kholy} in the form given by
\begin{eqnarray}\label{eprl+sf}
S[B,A,\Phi,\mu,\pi,\phi] =S_{GR}[B,A,\Phi,\mu] + \int_{\mathcal{M}^4}\left[ a \left(B_{IJ}\wedge \,^{\ast}B^{IJ}\right)
\pi^{\mu} \partial_{\mu}\phi + \left( \alpha_1 \h+\alpha_2 \g\right)\pi^{\mu}\pi^{\nu}d^4x\right ],
\end{eqnarray}
where $a$, $\alpha_1$ and $\alpha_2$ are constants and $\h$ and $\g$ are Urbantke metrics \cite{urbantke-m} of weight
one given by
\begin{subequations}\label{metricas}
 \begin{eqnarray}
\h&:=&\frac1{12}\n B_{\mu\alpha}^{IJ} B_{\beta\gamma}^{KL} B_{\delta\nu}^{MN}\eta_{JK}\eta_{LM}\eta_{NI},\\
\g&:=&\frac1{3}\n B_{\mu\alpha}^{IJ} B_{\beta\gamma}^{KL} B_{\delta\nu}^{MN}\eta_{IN}\varepsilon_{JMKL},
    \end{eqnarray}
\end{subequations}
where $B^{IJ}=\frac12 B^{IJ}_{\alpha\beta}\,dx^{\alpha}\wedge dx^{\beta}$. Here
$\tilde{\eta}^{\alpha\beta\gamma\delta}$ is such that $\tilde{\eta}^{0123}=1$; $\alpha$, $\beta$, $\gamma$,
$\ldots=0,1,2,3$ are spacetime indices, $d^4x= dx^0\wedge dx^1\wedge dx^2 \wedge dx^3$, and
$\varepsilon_{0123}=\epsilon$ equal to $+1$ or $-1$ depending on the orientation chosen.

The definition of the two metrics (\ref{metricas}) is inspired by the two Urbantke metrics introduced in Ref.
\onlinecite{cqgl2001}. Even though the metrics given in Eqs. (\ref{metricas}) are very closed to those reported in Ref.
\onlinecite{cqgl2001}, there is a slight difference between them: the ones reported here are covariant metrics of
weight one whereas the others are contravariant metrics of weight two\footnote{It is also possible to define
contravariant metrics of weight three given by $\tilde{B}^{\mu\alpha IJ}
\tilde{B}_{\alpha\beta}\,^{KL}\tilde{B}^{\beta\nu MN}\eta_{IN}\varepsilon_{JKLM}$ and $\tilde{B}^{\mu\alpha IJ}
\tilde{B}_{\alpha\beta}\,^{KL}\tilde{B}^{\beta\nu MN}\eta_{IN}\eta_{JK}\eta_{LM}$. These metrics and the ones of Ref.
\onlinecite{cqgl2001} could also be used to make the coupling of scalar field to general relativity or to other
theories of gravity by changing the tensorial nature and weight of the auxiliary fields involved. }. Notice also that
in Ref. \onlinecite{cqgl2001} were considered linear combinations of the metrics introduced therein. Similarly, the
Urbantke metrics used in Ref. \onlinecite{bimetric} are specific linear combinations of the metrics (\ref{metricas}).

The variation of this action with respect to the independent fields gives the equations of motion
\begin{subequations}\label{eq-sf}
\begin{eqnarray}
&& \delta B: \label{eq B} \left\{  (F_{IJKL} + \frac{1}{\gamma}{^*F}_{IJKL})e_{\gamma}^{K} e_{\delta}^{L}-
\Phi_{IJKL}B^{KL}_{\gamma\delta}+ a \varepsilon_{IJKL}
B^{KL}_{\gamma\delta} \phi^{\mu} \partial_{\mu}\phi \right. \nonumber\\
&&\hspace{4cm}\left.+\frac13 B_{\phi \gamma}^{KL} B_{\delta\theta}^{MN}\pi^{\phi}\pi^{\theta}\eta_{KN}
\left(\alpha_1\eta_{IL}\eta_{JM}+4\alpha_2\varepsilon_{IJLM}\right) \right\}
\tilde{\eta}^{\alpha\beta\gamma\delta} \nonumber \\
&&\hspace{4cm} +\frac23 B_{\phi\gamma}^{KL}B_{\delta\theta}^{MN}\pi^{\alpha}\pi^{\theta}\eta_{IN} \left(
\alpha_1\eta_{JK}\eta_{LM}+4\alpha_2\varepsilon_{JMKL}\right)\tilde{\eta}^{\beta\phi\gamma\delta}=0,\\
&& \delta A: D B^{IJ} + \frac{1}{\gamma} \,D\,^{\ast}B^{IJ} =0, \label{eq A}\\
&& \delta \Phi: B^{IJ} \wedge B^{KL} + 2 \mu\, \varepsilon^{IJKL}=0,\label{eq Phi}\\
&& \delta \mu:  \Phi_{IJKL} \varepsilon^{IJKL} =0, \label{eq mu}\\
&& \delta \pi: \left(B_{IJ}\wedge \,^{\ast}B^{IJ}\right)\,\partial_{\mu}\phi + 2 \left(\alpha_1 \h+\alpha_2 \g \right)
\pi^{\nu} d^4x=0, \label{eq pi}\\
&& \delta \phi: \partial_\mu\left(^{\ast}B^{IJ}_{\alpha\beta}\, B_{IJ\gamma\delta}\pi^\mu \n\right)=0. \label{eq phi}
\end{eqnarray}
\end{subequations}

Note that the equations of motion given in (\ref{eq A}) and (\ref{eq Phi}) are the same ones obtained in Sec. \ref{cc}
(see Eqs. (\ref{eq A CC}) and (\ref{eq Phi CC})). This means that the $B$ field has the expressions given in (\ref{sol
B's}) and that there is no torsion, so the curvature $F^{IJ}$ satisfies the Bianchi identities (\ref{bianchi}). In Sec.
\ref{cc} we used the equations of motion obtained from the proposed action principle (\ref{eprl+cc}) in order to show
that they imply the Einstein's equation with the cosmological constant. Additionally, it was also shown that the
Holst's action with the cosmological constant is obtained once the action principle (\ref{eprl+cc}) is written in terms
of the tetrad field. In this section we will focus our attention in rewriting the action principle (\ref{eprl+sf}) in
terms of the tetrads and the scalar field. In order to do this we need to solve (\ref{eq pi}) for the auxiliary field
$\pi$ in terms of the tetrad and the scalar field.

The substitution of the solutions given in (\ref{sol B's}) into (\ref{metricas}) and (\ref{eq pi}) will lead to two
different expressions for the field $\pi^\mu$ in terms of the tetrad field $e^I$. Let us analyze each case separately.

\subsection{Case $B^{IJ}=\kappa_1 \,{^{\ast}} \left ( e^I \wedge e^J \right )$}\label{sf-c1}
Inserting $B_{\alpha\beta}^{IJ}=\kappa_1 \varepsilon^{IJ}\,_{KL}e^K_\alpha\, e_{\beta}^L$ into Eq. (\ref{metricas})
yields to
\begin{subequations}\label{met c1}
 \begin{eqnarray}
\h&=&\kappa_1^3 \sigma\epsilon \dete g_{\mu\nu}[e], \\
\g&=&0.
\end{eqnarray}
\end{subequations}
where $g_{\mu\nu}[e]:= e^I_\mu e^J_\nu \eta_{IJ}$.

From (\ref{sol B's1}), (\ref{eq pi}), and (\ref{met c1}), we get
\begin{eqnarray}\label{pi c1}
\pi^\alpha=-\frac{6a}{\alpha_1\kappa_1} g^{\alpha\beta}[e]\,\partial_{\beta} \phi,
\end{eqnarray}
with $g^{\alpha\beta}[e]g_{\beta\gamma}[e]=\delta^{\alpha}_{\gamma}$.

Using (\ref{sol B's1}), (\ref{met c1}), (\ref{pi c1}), and $\dete=\sqrt{\sigma g}$ with
$g:=\det{(g_{\mu\nu}[e])}=(\dete)^2 \sigma$,  it is possible to rewrite the proposed action principle (\ref{eprl+sf})
as
\begin{eqnarray}\label{action e c1}
S[e,A,\phi]=\kappa_1 \int_{{\mathcal M}^4}\left\{\left[{^{\ast}} \left( e^I \wedge e^J \right) +\frac{\sigma}{\gamma}\,
e^I \wedge e^J \right]\wedge F_{IJ}[A]-8\pi G  \sqrt{\sigma g} g^{\mu\nu}[e] \partial_\mu \phi \,\partial_\nu \phi\,
d^4x\right\},\,\,
\end{eqnarray}
where it was considered $\frac{36 a^2}{\alpha_1}\, \sigma\epsilon=8 \pi G$ with $G$ the Newton constant. Now, it is
clear that $\frac{1}{16\pi G} $ times Eq. (\ref{action e c1}) is the wanted action in the first-order formalism, i.e.,
we have shown that Eq. (\ref{eprl+sf}) is equivalent to Holst's action for general relativity coupled to a scalar
field.

\subsection{Case $B^{IJ}=\kappa_2\,  e^I \wedge e^J $}\label{sf-c2}

When the two-form $B$ takes the form given in Eq. (\ref{sol B's2}), the Urbantke metrics (\ref{metricas}) acquire the
expressions
\begin{subequations}\label{met c2}
 \begin{eqnarray} \h&=&0, \\  \g&=&\kappa_2^3 \epsilon \dete g_{\mu\nu}[e].
 \end{eqnarray}
\end{subequations}
By plugging Eqs. (\ref{sol B's2}) and (\ref{met c2}) into (\ref{eq pi}) it is obtained that the field $\pi^\mu$ takes
the form
\begin{eqnarray}\label{pi c2}
\pi^\alpha=-\frac{6a}{\alpha_2\kappa_2} g^{\alpha\beta}[e]
\partial_{\beta} \phi.
\end{eqnarray}
So, in this case, the action (\ref{eprl+sf}) in terms of the tetrad field is given by
\begin{eqnarray}\label{action e c2}
S[e,A,\phi]=\frac{\kappa_2}{\gamma} \int_{{\mathcal M}^4}\left\{\left[{^{\ast}} \left( e^I \wedge e^J \right) +\gamma\,
e^I \wedge e^J \right]\wedge F_{IJ}[A]-8\pi G \epsilon \sqrt{\sigma g}\, g^{\mu\nu}[e]\partial_\mu \phi \,\partial_\nu
\phi\,d^4x\right\},\,\,
\end{eqnarray}
where it was used the relation $\frac{36 a^2}{\alpha_2}\gamma\epsilon = 8\pi G$.

\bigskip

Some remarks follow:
(i) By using the results obtained in Secs. \ref{sf-c1} and \ref{sf-c2}
, it is possible to fix the relationship among the constants $a$, $\alpha_1$, and $\alpha_2$, as
$\frac{\alpha_2}{\alpha_1}=\sigma \gamma$ and $\alpha_2=\frac{9 a^2 \gamma\epsilon}{2\pi G}$. Thus, we might define the
$BF$ action principle for gravity with scalar field as
\begin{eqnarray}
S[B,A,\Phi,\mu,\pi,\phi] &=&S_{GR}[B,A,\phi,\mu]  \nonumber \\
& &+ \int_{\mathcal{M}^4} \left [ a \left(B_{IJ}\wedge \,^{\ast}B^{IJ}\right) \pi^{\mu}
\partial_{\mu}\phi + \frac{9 a^2}{2\pi G} \left(\gamma \g + \sigma \h\right)
\pi^{\mu}\pi^{\nu}d^4x\right ]. \label{eprl+sf final}
\end{eqnarray}
As in the cases A and B, the equations of motion obtained from the variation of the action (\ref{eprl+sf final}) with
respect to the independent fields give the equations of general relativity with scalar field. Note that $a$ can be
absorbed by redefining
$\pi^{\mu}$.
(ii) Notice that we could have added the term $\left(B_{IJ}\wedge B^{IJ}\right) \pi^{\mu}
\partial_{\mu}\phi$ when we began the coupling of a scalar field in the action principle (\ref{eprl+sf}). Nevertheless, the inclusion of that term would
have followed a behavior similar to the one described in the last paragraph of Sec. \ref{cc} for the cosmological
constant.
(iii) We have shown that the action for gravity coupled to a scalar field in the first-order formalism has arisen from
the proposed $BF$ action (\ref{eprl+sf}). Therefore, the action (\ref{eprl+sf}) is a good action to describe the scalar
field coupling. Alternatively, the same conclusion can be reached by handling the equations of motion coming from the
action principle (\ref{eprl+sf}), i.e., the first four equations in (\ref{eq-sf}) become Einstein's equations coupled
to a scalar field whereas Eq. (\ref{eq phi}) becomes the Klein-Gordon equation once the auxiliary fields (\ref{pi c1})
or (\ref{pi c2}) are substituted into them.
(iv) Notice that the difference between the cases A and B previously found for the cosmological constant is also
present in the cases of the coupling of the scalar field.

\section{Coupling the Yang-Mills field}

In order to couple Yang-Mills fields to the action (\ref{eprl}) we consider
\begin{eqnarray}\label{eprl+YM}
S[B,A,\Phi,\mu,\mathcal{A},\phi]=S_{GR}[B,A,\Phi,\mu] + \int_{\mathcal{M}^4} K_{ab} \left( a\mathcal{F}^a[\mathcal{A}]
\wedge \phi^b_{IJ} B^{IJ} + b\mathcal{F}^a[\mathcal{A}] \wedge \phi^b_{IJ} {^{*}B}^{IJ}
\right.\hspace{2.5cm} \nonumber\\
- \left.\frac{\beta_1}2 \,\phi^a_{IJ}\phi^b_{KL} B^{IJ}\wedge B^{KL} -\frac{\beta_2}2 \,\phi^a_{IJ}\phi^b_{KL}
B^{IJ}\wedge {^{*}B}^{KL} \right),
\end{eqnarray}
where $a$, $b$, $\beta_1$ and $\beta_2$ are constants. The Yang-Mills field $\mathcal{A}=\mathcal{A}^a J_a$ and the
auxiliary field $\phi_{IJ}=\phi_{IJ}^a J_a$, with $[J_a,J_b]=f^c\,_{ab} J_c$, take values in the Lie algebra of the
gauge group and $K_{ab}$ is its Killing-Cartan metric. The variation of the action (\ref{eprl+YM}) with respect to the
independent fields gives
\begin{subequations}
 \begin{eqnarray}
&& \delta B: \label{eq B YM} F_{IJ}+\frac{1}{\gamma}{^*F}_{IJ}-\Phi_{IJKL}B^{KL}\nonumber\\ && \hspace{1.5cm}+\,
K_{ab}\left(a\ff^a \wedge \phi^b_{IJ}+b\ff^a \wedge
{^*\phi}^b_{IJ}-\beta_1\phi^a_{IJ}\phi^b_{KL}B^{KL}-\frac{\beta_2}{2}\phi^a_{IJ} {^*\phi}^b_{KL} B^{KL}
-\frac{\beta_2}{2}\phi^a_{KL} {^*\phi}^b_{IJ} B^{KL}\right)=0,
\\
&& \delta A: D B^{IJ} + \frac{1}{\gamma} \,D\,^{\ast}B^{IJ} =0, \label{eq A YM}\\
&& \delta \Phi: B^{IJ} \wedge B^{KL} + 2 \mu\, \varepsilon^{IJKL}=0,\label{eq Phi YM}\\
&& \delta \mu:  \Phi_{IJKL} \varepsilon^{IJKL} =0, \label{eq mu YM}\\
&& \delta \mathcal{A}^a: \mathcal{D}(\phi_{aIJ}B^{IJ})=0, \label{eq Atilde YM}\\
&& \delta \phi^a: K_{ab} \left( a\mathcal{F}^a\wedge B^{IJ} + b\mathcal{F}^a\wedge {^{*}B}^{IJ}- \beta_1 \,\phi^a_{KL}
B^{IJ}\wedge B^{KL} -\frac{\beta_2}2 \,\phi^a_{KL} B^{KL}\wedge ^{*}B^{IJ} -\frac{\beta_2}2 \,\phi^a_{KL} B^{IJ}\wedge
^{*}B^{KL}\right)=0,\,\, \nonumber\\ \label{eq phi YM}
 \end{eqnarray}
\end{subequations} where, it was used the notation $\mathcal{F}^a$ for $\mathcal{F}^a[\mathcal{A}]=d \mathcal{A}^a+\frac12
f^a\,_{bc}\mathcal{A}^b \wedge\mathcal{A}^c$, and the definition $\mathcal{D}u^a=d u^a+f^{a}\,_{bc} \mathcal{A}^b
\wedge u^c$ for $u=u^a J_a$.

Notice that, since Eqs. (\ref{eq A YM}) and (\ref{eq Phi YM}) are equal to (\ref{eq A}) and (\ref{eq Phi}), the
modified action (\ref{eprl+YM}) keeps two basic properties: $A$ is the spin-connection (thus its curvature satisfies
(\ref{bianchi})) and the two-form $B$'s can be written in terms of the tetrad field as given in (\ref{sol B's}). We are
going to show that the proposed action (\ref{eprl+YM}) is equivalent to Holst's action for gravity coupled to the
Yang-Mills field. As it should be clear to the reader by now, the simplest way to obtain the goal is to solve the Eq.
(\ref{eq phi YM}) for the auxiliary fields. Again, it is convenient to consider each case of (\ref{sol B's})
separately.

\subsection{Case $B^{IJ}=\kappa_1 \,{^{\ast}} \left ( e^I \wedge e^J \right )$}\label{YM1}

Taking for $B^{IJ}$ the expression given in (\ref{sol B's1}) and assuming $\dete \neq 0$ it is possible to solve
(\ref{eq phi YM}) for $\phi^a_{IJ}$ as
\begin{eqnarray}\label{phi c1 YM}
\phi^{a}_{IJ}=\frac{\sigma}{2\kappa_1}\A \left[(a\beta_2-b\beta_1) \ff^a_{IJ}+(b \beta_2-\sigma a \beta_1)
^{\ast}\ff_{IJ}^a\right],
\end{eqnarray}
with $^{\ast}\ff_{IJ}^a=\frac12 \varepsilon_{IJ}\,^{KL}\ff_{KL}^a$. By plugging this expression and the solution
(\ref{sol B's1}) into the action (\ref{eprl+YM}), and using the Bianchi identities (\ref{bianchi}), the action
(\ref{eprl+YM}) becomes
\begin{eqnarray}\nonumber
S[e,\mathcal{A}] = \kappa_1 \int_{{\mathcal M}^4} \left[ \right. \epsilon \dete F^{IJ}\,_{IJ}[A[e]] d^4x \hspace{10cm}
\\ \left.+\,\frac{K_{ab}}{2\kappa_1} \A  \left\{[2ab\beta_2-(a^2\sigma+b^2)\beta_1] \ff^a\wedge \ff^b + [(a^2
\sigma+b^2)\beta_2-2ab\beta_1\sigma]\ff^a\wedge {^*\ff}^b\right\} \right],\label{YM c1}
\end{eqnarray}
which is the usual action that describes the coupling of the Yang-Mills field to general relativity in the first-order
formalism supplemented with the Pontrjagin term.

\subsection{Case $B^{IJ}=\kappa_2\,  e^I \wedge e^J $}

Following an analogous procedure to the case A, i.e. considering (\ref{sol B's2}) and assuming $\dete \neq 0$, Eq.
(\ref{eq phi YM}) can be solved for $\phi^a_{IJ}$ as
\begin{eqnarray}\label{phi c2 YM}
\phi^{a}_{IJ}=\frac{\sigma}{2\kappa_2}\A \left[(b\beta_2\sigma-a\beta_1)\ff^a_{IJ}+(a\beta_2- b\beta_1)\,
^{\ast}\ff_{IJ}^a\right],
\end{eqnarray}
and by using this solution the action (\ref{eprl+YM}) acquires the form
\begin{eqnarray}\nonumber
S[e,\mathcal{A}] \label{YM c2} =\frac{\kappa_2}{\gamma}\int_{{\mathcal M}^4} \left[\epsilon\dete F^{IJ}\,_{IJ}[A[e]]
d^4x \hspace{10cm} \right.
\\ \left.+ \,\frac{\gamma\sigma K_{ab}}{2\kappa_2} \A \left( [2ab\beta_2-(a^2\sigma+b^2)\beta_1]\ff^a\wedge \ff^b+[ (a^2
\sigma+b^2)\beta_2-2ab\beta_1\sigma] \ff^a\wedge {^*\ff}^b \right)\right].
\end{eqnarray}

\medskip
We conclude this section by making some remarks:
(i) Note that the coefficients of the Pontrjagin and Yang-Mills terms
that appear in the final action principles (\ref{YM c1}) and (\ref{YM c2}) depend on the whole set of coefficients of
the terms added to the original action principle in (\ref{eprl+YM}). Observe also that different combinations of those
terms allow us to make the coupling of the Yang-Mills field into the action principle (\ref{eprl}).
(ii) Notice that a term of the form $\frac{\beta_3}2 \,\phi^a_{IJ}\phi^b_{KL} {^{\ast}B}^{IJ}\wedge {^{\ast}B}^{KL}$
can be added to the action principle (\ref{eprl+YM}). If such a term is included its effect would be a redefinition of
the $\beta_1$ parameter.
(iii) Finally, note that the coupling of the Yang-Mills field is also useful to obtain the coupling of the Maxwell field by
doing the replacement $\phi^a_{IJ}\rightarrow \phi_{IJ}$ in the auxiliary fields.

\section{Concluding remarks}

We have done the coupling of the cosmological constant, the scalar, Yang-Mills, and Maxwell fields to general
relativity in the theoretical framework of $BF$ gravity with Immirzi parameter. The analysis was carried out using the
action principle for gravity used in Ref. \onlinecite{epr}. Our analysis shows that the coupling can be done without
any technical or conceptual difficulties. Therefore, the framework of $BF$ gravity with Immirzi parameter is robust
enough to allow the coupling of matter fields. The action principles for the matter fields include all the terms that
{\it a priori} might contribute to the coupling. The role that each one of these terms plays can be clearly appreciated
when the action principles are written in the first-order and second-order formalisms. This fact also allows us to see
that the proposed action principles for the coupling of the matter fields are the right ones. From these action
principles, Einstein's equations and matter field equations follow immediately. This was explicitly shown for the
coupling of the cosmological constant. For the scalar field, Maxwell, and Yang-Mills field it was shown that the
proposed action principles are equivalent to the usual action principles in the first-order formalism.

Notice that the results presented in this work can be used to obtain the coupling of matter fields to the CMPR action
principle by performing the transformation introduced in Ref. \onlinecite{sigma1}. It could be interesting to compare
the resulting couplings with the ones obtained directly from the CMPR action principle (see Refs. \onlinecite{prd2010}
and \onlinecite{barbitas}). For example, in the case of the cosmological constant these two approaches agree as can be
seen from the comparison of the results presented in the Sec. \ref{cc} and the ones presented in the Sec. 4 of Ref.
\onlinecite{sigma1} where the coupling of the cosmological constant to the action principle (\ref{eprl}) is obtained by
doing a transformation from the CMPR action principle.

We think that our approach and results are relevant because they display the way matter fields couple to general
relativity written as a constrained $BF$ theory. The couplings are restricted by the geometric and tensor nature of the
$B$ fields, the matter fields and the auxiliary fields themselves. Therefore, when all these fields are put together,
they lead to the coupling terms introduced in this paper. The simultaneous coupling of all the matter fields is
immediate. Even though our results are classical they might be useful or interesting for the spinfoam approach to
quantizing gravity because it is based on a reformulation of general relativity as a constrained $BF$ theory
\cite{spinfoam}.

Our results are also relevant for the modification of general relativity recently proposed by Krasnov
\cite{kr1,kr2,kr3} (and discussed in Refs. \onlinecite{mpvq2}, \onlinecite{bimetric}, \onlinecite{freidel},
\onlinecite{cqg2010}, \onlinecite{diego}, and \onlinecite{diego2}) aiming to developing a renormalizable theory of
gravity from $BF$ theory. This is so because the matter coupling terms introduced in this paper can be used to make the
matter couplings in the framework of such theories.

The coupling of fermion fields of spin $\frac12$, $\frac32$, etc. to $BF$ gravity was not analyzed in this paper. The
reason is that such a coupling deserves a separate treatment because of the various types of fermions and also because
of the various ways fermions couple to gravity, but such an analysis must be done and confronted to the classical and
semiclassical limits of the quantum theory developed in Ref. \onlinecite{fermiones}. This issue as well as the study of
supersymmetric fields and exotic matter is left for future work.


\section*{ACKNOWLEDGMENTS}
These results were presented in the International Conference on Quantum Gravity LOOPS11 held in Madrid, Spain, 2011.
Warm thanks to Riccardo Capovilla and Alejandro Corichi for very fruitful discussions on the subject of this paper.
This work was supported in part by CONACYT, Mexico, Grant No. 167477-F.



\begin{thebibliography}{100}

\bibitem{lqg}
    T. Thiemann, {\it Modern Canonical Quantum General Relativity} (Cambridge University Press, Cambridge, England, 2007);
    C. Rovelli, {\it Quantum Gravity} (Cambridge University Press, Cambridge, MA 2004);
    A. Ashtekar and J. Lewandowski, {\it Class. Quantum Grav.} {\bf 21}, R53 (2004);
    A. Perez, 
        Proceedings of the International Conference on Fundamental Interactions,
        Domingos Martins, Brazil, (2004), arXiv:gr-qc/0409061;
    A. Ashtekar, {\it Lectures on Non-Perturbative Canonical Gravity} (World Scientific, Singapore, 1991);
    C. Rovelli, 
        {\it Living Rev. Relativity} {\bf 1}, 1 (1998);
    A. Ashtekar and B. Krishnan, 
        {\it Living Rev. Relativity} {\bf 7}, 10 (2004);
    M. Bojowald, 
        {\it Living Rev. Relativity} {\bf 11}, 4 (2008);
    M. Gaul and C. Rovelli, {\it Lect. Notes Phys.} {\bf 541}, 277 (2000);
    L. Smolin, ``An invitation to loop quantum gravity'', arXiv:hep-th/0408048;
    A. Ashtekar, {\it Nuovo Cimento B} {\bf 122}, 135 (2007);
    C. Rovelli, {\it Class. Quantum Grav.} {\bf 28}, 114005 (2011).
\bibitem{spinfoam}
    A. Perez, {\it Class. Quantum Grav.} {\bf 20}, R43 (2003);
    D. Oriti, {\it Rept. Prog. Phys.} {\bf 64}, 1703 (2001);
    J.C. Baez, {\it Lect. Notes Phys.} {\bf 543}, 25 (2000); {\it Class. Quantum Grav.} {\bf 15}, 1827 (1998);
    J. Engle, R. Pereira, and C. Rovelli, {\it Phys. Rev. Lett.} {\bf 99}, 161301 (2007);
    E. R. Livine and S. Speziale, {\it Europhys. Lett.} {\bf 81}, 50004 (2008);
    C. Rovelli, 
        {\it Living Rev. Relativity} {\bf 11}, 5 (2008);
    S. Alexandrov and P. Roche, {\it Phys. Rep.} {\bf 506}, 41 (2011);
    E. Livine, ``The Spinfoam Framework for Quantum Gravity'', arXiv:1101.5061v1[gr-qc];
    A. Perez,  {\it Living Rev. Relativity} (unpublished).
\bibitem{vlad1} V. Cuesta and M. Montesinos, {\it Phys. Rev.} D {\bf 76}, 104004
    (2007).
\bibitem{vlad2} V. Cuesta, M. Montesinos, M. Vel\'azquez, and J.D. Vergara,
    {\it Phys. Rev.} D {\bf 78}, 064046 (2008); arXiv:0809.2741.
\bibitem{meche1} M. Montesinos and M. Vel\'azquez, in {\it The Planck
    Scale, XXV Max Born Symposium}, edited by J. Kowalski-Glikman, R. Durka, and
    M. Szczachor, AIP Conference Proceedings {\bf 1196} (American Institute
    of Physics, New York, 2009) p. 201; arXiv:0812.2825.
\bibitem{meche2} M. Montesinos and M. Vel\'azquez, in {\it The Twelfth Marcel Grossmann Meeting on
    General Relativity: Recent Developments in Theoretical and Experimental General Relativity,
    Astrophysics, and Relativistic Field Theories}, edited by T. Damour, R. Jantzen, and R. Ruffini (World Scientific, Singapore,
    2011) p. 2294.
\bibitem{cuando} M. Montesinos, A. Perez, and M. Vel\'azquez (work in
    progress).
\bibitem{pleb77}
    J.F. Pleba\'nski, {\it J. Math. Phys. (N.Y.)} {\bf 18}, 2511 (1977).
\bibitem{1Samuel}
    J. Samuel, {\it Pramana J. Phys.} {\bf 28}, L429 (1987).
\bibitem{1Jacobson}
    T. Jacobson and L. Smolin, {\it Phys. Lett. B} {\bf 196},
    39 (1987); {\it Class. Quantum Grav.} {\bf 5}, 583 (1988).
\bibitem{cdjm}
    R. Capovilla, J. Dell, T. Jacobson, and L. Mason, {\it Class. Quantum Grav.} {\bf 8}, 41 (1991).
\bibitem{cqgl2001}
    R. Capovilla, M. Montesinos, V.A. Prieto, and E. Rojas,
    {\it Class. Quantum Grav.} {\bf 18}, L49 (2001).
\bibitem{epr}
    J. Engle, R. Pereira, and C. Rovelli, {\it Nucl. Phys.} B {\bf 798}, 251 (2008).
\bibitem{sigma1}
    M. Montesinos and M. Vel\'azquez, 
    {\it SIGMA} {\bf 7}, 103 (2011); arXiv:1111.2671[gr-qc].
\bibitem{oriti}
    R.E. Livine and D. Oriti, {\it Phys. Rev.} D {\bf 65}, 044025 (2002).
\bibitem{prd2010}
    M. Montesinos and M. Vel\'azquez, {\it Phys. Rev.} D {\bf 81}, 044033 (2010); arXiv:1002.3836 [gr-qc].
\bibitem{mpvq2}
    M. Vel\'azquez, {\it Ph.D. thesis}, Cinvestav, Mexico, 2011.
\bibitem{celada}
    M. Celada, {\it M.Sc. thesis}, Cinvestav, Mexico, 2011.
\bibitem{celada2}
    M.A. Celada Mart\'{\i}nez, {\it An\'alisis hamiltoniano de gravedad a la BF} (Editorial Acad\'emica Espa\~nola,
    Buenos Aires, 2012). 
\bibitem{simone}
    L. Smolin and S. Speziale,
    {\it Phys. Rev. D} {\bf 81}, 024032 (2010). 
\bibitem{merced}
    M. Montesinos, {\it Class. Quantum Grav.} {\bf 23}, 2267 (2006).
\bibitem{kholy}
    A.A. El-Kholy, R.U. Sexl, and H.K. Urbantke, {\it Ann. Inst. Henri Poincar\'e} (A) {\bf 18}, 121 (1973).
\bibitem{urbantke-m}
    H. Urbantke, {\it J. Math. Phys. (N.Y.)} {\bf 25}, 2321 (1984).
\bibitem{bimetric}
    S. Speziale, {\it Phys. Rev.} D {\bf 82}, 064003 (2010).
\bibitem{barbitas}
    B.A. Ju\'arez Aubry, {\it B.Sc. thesis}, Fundaci\'on Universidad de las Am\'ericas Puebla, Mexico, 2011.
\bibitem{kr1} K. Krasnov, ``Renormalizable Non-Metric Quantum Gravity?''; arXiv:hep-th/0611182.
\bibitem{kr2} K. Krasnov, {\it Mod. Phys. Lett.} A {\bf 22}, 3013 (2007).
\bibitem{kr3} K. Krasnov, {\it Phys. Rev. Lett.} {\bf 100}, 081102 (2008).
\bibitem{freidel}
    L. Freidel, ``Modified gravity without new degrees of freedom''; arXiv:0812.3200[gr-qc].
\bibitem{cqg2010}
    R. Capovilla, M. Montesinos, and M. Vel\'azquez, {\it Class. Quantum Grav.} {\bf 27}, 145011 (2010).
\bibitem{diego}
    D. Gonz\'alez, {\it M.Sc. thesis}, Cinvestav, Mexico, 2011.
\bibitem{diego2}
    D.G. Gonz\'alez, {\it Gravedad de 2-formas} (Editorial Acad\'emica Espa\~nola, Buenos Aires, 2012). 
\bibitem{fermiones}
    E. Bianchi, M. Han, E. Magliaro, C. Perini, C. Rovelli, and W. Wieland, ``Spinfoam fermions'',
    arXiv:1012.4719v1[gr-qc].
\end{thebibliography}
\end{document}